\documentclass [pra,twocolumn,showpacs]{revtex4}
\usepackage{dcolumn,bm}
\newcommand{\half}{\ensuremath{\frac{1}{2}}}
\newcommand{\eref}[1]{(\ref{#1})}
\begin{document}
\title{Calculation of nuclear spin-dependent parity-nonconserving
amplitude for $(7s,F=4) \rightarrow (7s,F=5)$ transition in Fr}
\author{S.~G.~Porsev}
\email{porsev@thd.pnpi.spb.ru}
\author{M.~G.~Kozlov}
\affiliation{Petersburg Nuclear Physics Institute, Gatchina,
Leningrad district, 188300, Russia}
\date{\today}

\begin{abstract}
Many-body calculation of nuclear spin-dependent
parity-nonconserving amplitude for $7s,F=4 \rightarrow 7s,F=5$
transition between hyperfine sublevels of the ground state of
$^{211}$Fr is carried out. The final result is $ \langle 7s,F=5 ||
d_{\rm PNC}|| 7s,F=4 \rangle = -0.49
\times 10^{-10} \, i \, \kappa \; {\rm a.u.},$ where $\kappa$ is
the dimensionless coupling constant. This is approximately an order
of magnitude larger than similar amplitude in Cs. The dominant
contribution to $\kappa$ is associated with the anapole moment of
the nucleus.
\end{abstract}
\pacs{32.80Ys, 11.30.Er, 31.30.Jv}
\maketitle

\paragraph{Introduction.}
In this work we calculated nuclear spin-dependent
parity-nonconserving (PNC) amplitude for $7s,F=4 \rightarrow
7s,F=5$ transition between hyperfine structure components of the
ground state of the odd isotope of francium $^{211}$Fr. Three
effects contribute to this amplitude \cite{Khr91}: the interaction
of an electron with the nuclear anapole moment (AM), the
electron-nuclear neutral current interaction, and the combined
action of the nuclear spin-independent electron-nucleus weak
interaction and the hyperfine interaction.

The AM $\bm a$ was introduced by Zel'dovich \cite{Zel57t} just
after the discovery of parity violation. A first realistic model
for the AM of the nucleus was suggested in
Refs.~\cite{FK80,FKS84a}. There it was shown that for heavy nuclei
$a\sim A^{2/3}$, where $A$ is the number of nucleons. AMs of the
nuclei with unpaired proton are expected to be few times larger
than for the case of unpaired neutron. Because of that for atoms
with large and odd $Z$ the AM contribution to the spin-dependent
part of the PNC amplitudes dominates over that of the
electron-nucleon neutral currents. The third contribution is also
$\sim A^{2/3}$, but is numerically smaller \cite{FK85a,BP91b} (see
Eq.~\eref{e1b} below). Note that the neutral current and hyperfine
contributions to the nuclear spin-dependent PNC amplitude are well
known from the standard model. Therefore, any measurement of the
respective coupling constant $\kappa$ will give unambiguous
information about AM of the nucleus.

For the optical transitions in heavy atoms the spin-independent
PNC amplitudes are approximately two orders of magnitude larger
than the spin-dependent ones. Because of that the AM was measured
experimentally only for cesium \cite{WBC97}. This measurement
provided a valuable probe of the relatively poorly understood
parity nonconservation in nuclei \cite{FM97,HW01}. Further
experimental and theoretical investigations of AM are very
important both for nuclear physics and for physics of the
fundamental interactions.

An alternative possibility to observe the spin-dependent PNC
amplitudes was suggested in \cite{NK75t}: in the rf transitions
between the hyperfine components of the ground state of an atom
the spin-independent amplitudes are negligible and the dominant
PNC effect is caused by the AM. Using the rf resonator one can
have an additional enhancement of the PNC effect by placing the
gas cell in the node of the magnetic and the antinode of the
electric rf fields \cite{GEKM88}. The PNC effect can be also
enhanced in the strong dc magnetic field \cite{Fla88}. The new
cooling and trapping techniques allow to increase the intensity of
the rf transitions making these experimental schemes much more
realistic. At present there is an ongoing project of measuring PNC
effects in francium \cite{Beh93,Oro97} and the observation of the
AM in the hyperfine transition can be a valuable addition to this
project.

Semi-empirical calculations of nuclear spin-dependent amplitudes
for transitions between hyperfine sublevels of the ground state
were already carried out for Cs and Tl \cite{NK75t} and for K and Cs
\cite{GEKM88}. Fr is the heaviest of alkali-metal atoms. Since
spin-dependent amplitude grows with nuclear charge $Z$ faster than
$Z^2$, one can expect that for Fr this amplitude will be
significantly larger than for other alkali-metal atoms. Besides
that, a large number of odd isotopes with non-zero nuclear spin
makes it possible (at least in principle) to study dependence of
nuclear spin-dependent amplitude on the nuclear structure.

\paragraph{Theory.}
It is known that parity-nonconserving electron-nuclear interaction
can be divided into two parts: nuclear spin-independent part and
nuclear spin-dependent one. The respective PNC Hamiltonian can be
written as follows (atomic units are used throughout the paper):
\begin{eqnarray}
     H_{\rm PNC} = H_{\rm SI} + H_{\rm SD}
      = \frac{G_F}{\sqrt{2}}                             
     \Bigl(-\frac{Q_W}{2} \gamma_5 + \frac{\kappa}{I}
     {\bm \alpha} {\bm I} \Bigr) \rho({\bm r}),
\label{e1}
\end{eqnarray}
where  $G_F = 2.2225 \times 10^{-14}$ a.u. is the Fermi constant of
the weak interaction, $Q_W$ is the nuclear weak charge, $\kappa$ is
the dimensionless coupling constant, $\bm \alpha =
\gamma_0 {\bm \gamma}$, $\gamma_i$ are the Dirac matrices, $\bm I$ is the
nuclear spin ($I=\frac{9}{2}$ for the isotope $^{211}$Fr), and $\rho({\bm
r})$ is the nuclear density distribution.

As we mentioned above, there are three main contributions to the
coupling constant $\kappa$ in the spin-dependent part of the PNC
interaction \eref{e1}:
\begin{eqnarray}
     \kappa &=& (-1)^{I+\half-l}\frac{I+\half}{I+1}\kappa_a
     +\kappa_2 + \kappa_{Q_W},                                 
\label{e1a}
\end{eqnarray}
where the anapole contribution is given by the constant $\kappa_a$
\cite{FK80} ($l$ is the orbital angular momentum of the unpaired
nucleon), the constant $\kappa_2$ corresponds to the spin-dependent
weak neutral currents and the term $\kappa_{Q_W}$ is induced by the
interference of the spin-independent PNC interaction with the
hyperfine interaction. For heavy nuclei constants $\kappa_a$ and
$\kappa_{Q_W}$ are proportional to $A^{2/3}$ \cite{FK85a,BP91b},
and their ratio depends on the dimensionless constant of the weak
interaction of the unpaired nucleon with the nuclear core $g$
\cite{FM97}:
\begin{eqnarray}
     \frac{\kappa_{Q_W}}{\kappa_a}
     &\approx& q \frac{N \mu_N}{A \mu g} ,                 
\label{e1b}
\end{eqnarray}
where $\mu_N$ and $\mu$ are magnetic moments of the nucleus and
the valence nucleon, correspondingly. The numerical factor $q$ is
within the limits $1 < q < 3$ (see, e.~g., \cite{FM97}). For the
unpaired proton $g_p \approx 7$, while for neutron $g_n \approx
-2$ \cite{FM97} (see also \cite{HW01}). This estimate shows that
for the odd isotopes of Fr the anapole contribution dominates in
\eref{e1a}.

We assume that the nucleus is a uniformly charged sphere:
$$\rho({\bm r}) = \frac{3}{4 \pi r^3_n} \Theta (r_n-r).$$
The root-mean-square charge radius for $^{211}$Fr was measured to
be $r_{\rm rms}$ = 5.566 fm \cite{GOP99}. Using the relation
$ r_n = \sqrt{\frac{5}{3}} \ r_{\rm rms},$
we find $r_n$ = 7.186 fm.

If $|i \rangle$ and $|f \rangle$ are initial and final atomic
states of the same nominal parity, then to the lowest nonvanishing
order, the electric dipole transition matrix element (ME) is equal
to:
\begin{eqnarray}
   \langle f | d_{q,\rm PNC}  | i \rangle  &=&  \sum_{n} \left[      
\frac{\langle f | d_q | n  \rangle
      \langle n | H_{\rm PNC} | i \rangle}{E_i - E_n}\right.
\nonumber \\
      &+&
\left.\frac{\langle f | H_{\rm PNC} | n  \rangle
      \langle n | d_q | i \rangle}{E_f - E_n} \right],
\label{e2}
\end{eqnarray}
where $|a \rangle \equiv |J_a,F_a,M_a \rangle$ and ${\bm F} = {\bm
I} + {\bm J}$ is the total angular momentum.

In our case the contribution of $H_{\rm SI}$ (see Eq.~(\ref{e1}))
is negligible, so we consider only the nuclear spin-dependent part
of the PNC Hamiltonian. The ME of $H_{\rm SD}$ can be written as
follows
\begin{eqnarray}
  \langle n |H_{\rm SD}| i \rangle & = &
  (-1)^{I+F_i+J_i} \sqrt{I(I+1)(2I+1)}
    \delta_{F_n F_i} \delta_{M_n M_i}         \nonumber \\
 &\times&  \left\{ \begin{array}{ccc}
     J_n & J_i &  1   \\
      I  &  I  & F_i   \\                                     
     \end{array} \right\}
    \langle J_n || H_{\rm SD} || J_i \rangle \, ,
\label{e3}
\end{eqnarray}
where
$ \langle J_n || H_{\rm SD} || J_i \rangle =
\frac{G_F}{\sqrt{2}} \frac{\kappa}{I}
   \langle J_n ||\gamma_0\gamma \rho({\bm r})|| J_i \rangle. $

The ME of the operator $d_q$ is given by the following expression:
\begin{widetext}
\begin{eqnarray}
    \langle f | d_q | n \rangle &=&
    (-1)^{F_f-M_f} \left( \begin{array}{ccc}
                           F_f & 1 & F_n  \\
                          -M_f & q & M_n   \\
                           \end{array} \right)
    (-1)^{I+F_n+J_f+1} \sqrt{(2F_n+1)(2F_f+1)}
\left\{ \begin{array}{ccc}
     J_n & J_f & 1   \\
     F_f & F_n & I   \\                                  
     \end{array} \right\}
 \langle J_f || d || J_n \rangle.
\label{e4}
\end{eqnarray}
Applying the Wigner-Eckart theorem to the PNC amplitude:
$$ \langle f | d_{q,\rm PNC} | i \rangle =
      (-1)^{F_f-M_f} \left( \begin{array}{ccc}
                           F_f & 1 & F_i  \\
                          -M_f & q & M_i   \\
                           \end{array} \right)
   \langle J_f,F_f || d_{\rm PNC} || J_i,F_i \rangle, $$
and substituting Eqs.~(\ref{e3}) and~(\ref{e4}) in Eq.~(\ref{e2})
we get the following expression for the reduced ME of the PNC
amplitude:
\begin{eqnarray}
\label{e5}
     &&\langle J_f,F_f || d_{\rm PNC} || J_i,F_i \rangle =
     \sqrt{I(I+1)(2I+1)(2F_i+1)(2F_f+1)} 
     \sum_{n} \left[ (-1)^{J_f - J_i}
     \left\{ \begin{array}{ccc}
     J_n  &  J_i  &   1    \\
      I   &   I   &  F_i   \\                                    
     \end{array} \right\}
     \left\{ \begin{array}{ccc}
      J_n  &  J_f  &  1   \\
      F_f  &  F_i  &  I   \\
     \end{array} \right\} \right.
\\
  &&\times \frac{ \langle J_f || d || n,J_n \rangle
     \langle n,J_n || H_{\rm SD} || J_i \rangle }{E_n - E_i} \, +
     (-1)^{F_f - F_i}
     \left\{ \begin{array}{ccc}
     J_n  &  J_f  &   1    \\
      I   &   I   &  F_f   \\
     \end{array} \right\}
     \left\{ \begin{array}{ccc}
     J_n  &  J_i  &  1   \\
     F_i  &  F_f  &  I   \\
     \end{array} \right\}
     \left. \frac{\langle J_f || H_{\rm SD} ||n,J_n \rangle
            \langle n,J_n || d ||J_i \rangle}{E_n - E_f}  \right].
\nonumber
\end{eqnarray}
Note that for the transition between the hyperfine components of
the ground state $7s_{\half}$, one has $J_i=J_f=\half$,
$E_i=E_f=E_{7s}$, and $F_i=F_f-1=I-\half$. That leads to some
simplification of Eq.~(\ref{e5}):
\begin{eqnarray}
\label{e5a}
     \langle 7s_{\half},F_f || d_{\rm PNC} || 7s_{\half},F_i \rangle
     &=& 2I(I+1)\sqrt{(2I+1)}
     \sum_{n} \frac{ \langle 7s_{\half} || d || n,J_n \rangle
     \langle n,J_n || H_{\rm SD} || 7s_{\half} \rangle }{E_n - E_{7s}}
\\&\times&     \left[
     \left\{ \begin{array}{ccc}
         I  &  I  &   1    \\
      \half & J_n &  F_i   \\
     \end{array} \right\}
     \left\{ \begin{array}{ccc}
        F_i  &  J_n  &  I   \\
      \half  &  F_f  &  1   \\
     \end{array} \right\}                                        
+
     \left\{ \begin{array}{ccc}
         I  &  I  &   1    \\
      \half & J_n &  F_f   \\
     \end{array} \right\}
     \left\{ \begin{array}{ccc}
        F_f  &  J_n  &  I   \\
      \half  &  F_i  &  1   \\
     \end{array} \right\}
  \right],
\nonumber
\end{eqnarray}
\end{widetext}
where the sum runs over the states of odd-parity with angular
momenta $J_n=\half,\frac{3}{2}$. Novikov and Khriplovich pointed
out, that for alkali atoms contribution of the intermediate states
with $J_n=\frac{3}{2}$ is strongly suppressed. If these states are
excluded from the sum in Eq.~(\ref{e5a}), it can be further
simplified to the form:
\begin{eqnarray}
\label{e5b}
     &&\langle 7s_{\half},F_f || d_{\rm PNC} || 7s_{\half},F_i \rangle
     = \frac{2}{3} \sqrt{I(I+1)(I+\half)}
\nonumber
\\ &&\times
     \sum_{n} \frac{ \langle 7s_{\half} || d || n,\half \rangle
     \langle n,\half || H_{\rm SD} || 7s_{\half} \rangle }
     {E_n-E_{7s}}.
\end{eqnarray}
Eq.~(\ref{e5b}) was used in the semiempirical calculations
\cite{NK75t,GEKM88}, but here we use the more accurate expression
(\ref{e5a}).

\paragraph{Method of calculations and results.}

The Dirac-Fock-Breit equations were solved self-consistently on a
radial grid for the core electrons [1$s$,...,6$p_{\frac{3}{2}}$].
Then, the valence orbitals $7s,7p,8s,8p,9p$ were constructed in
V$^{N-1}$ approximation. The basis set used in calculations
included also virtual orbitals up to $25s, 25p, 24d, 15f$, and
$15g$ formed with the help of the method described in
Refs.~\cite{Bv83,Bog91,KPF96,KP97}.

To find the nuclear spin-dependent PNC amplitude defined by
Eq.~(\ref{e5a}), one needs to sum over intermediate states or
solve the corresponding inhomogeneous equations (Sternheimer
\cite{Ste50} or Dalgarno-Lewis \cite{DL55} method). Here we apply
the Sternheimer-Dalgarno-Lewis method to the valence part of the
problem as described in~\cite{KPF96,KP99}. Solving inhomogeneous
equation we find the answer in the Dirac-Fock approximation:
\begin{equation}
    \langle 7s, \ F=5 || d_{\rm PNC} || 7s, \ F=4 \rangle_{\rm DF}
    = -0.42 \times 10^{-10} \, i \, \kappa \; {\rm a.u.}.
\label{s1}
\end{equation}

It is known that core-valence correlations usually play an
important role for the PNC amplitudes. We first solved the
random-phase approximation (RPA) equations, summing a certain
sequence of many-body diagrams to all orders for both operators in
the right hand side of Eq.~\eref{e2}. Note that after the RPA
equations are solved for the operator $H_{\rm SD}$, the MEs
$\langle ns ||H_{\rm SD}|| np_{\frac{3}{2}} \rangle$ are no longer
equal to zero. As a result, the intermediate $np_{\frac{3}{2}}$
states also contribute to the spin-dependent PNC amplitude. We
found that their contribution to $\langle 7s,F=5 || d_{\rm PNC}||
7s,F=4 \rangle$ is about 10\%. That contribution is neglected in
the approximation (\ref{e5b}). The RPA correction changes the PNC
amplitude to:
\begin{equation}
   \langle 7s, \ F=5 || d_{\rm PNC} || 7s, \ F=4 \rangle_{\rm RPA}
   = -0.48 \times 10^{-10} \, i \, \kappa \; {\rm a.u.}.
\label{s2}
\end{equation}

The core polarization was taken into account by many-body perturbation
theory. We completely accounted for the second order of perturbation
theory and partly for the higher orders. In particular, we calculated
the structural radiation and normalization corrections to the PNC
amplitude.

Finally, taking into account that the initial and final states are
the many-electron states one needs to account for the excitations
from the $np_j$ shells ($n$=2--6). Respective ``core'' contribution
to the spin-dependent amplitude was estimated to be $-$3.5\%

\begin{table}
\caption{Nuclear spin-dependent PNC amplitude $\langle 7s,F=5 ||
d_{\rm PNC}|| 7s,F=4 \rangle$ in the units $i \times 10^{-10}
\kappa$. The 1-st column present the result obtained in the
Dirac-Fock approximation for the Coulomb-Gaunt potential.
Following columns present corrections discussed in the text. In
the column MBPT the Brueckner, structural radiation and
normalization corrections are summed together. In the column
``core'' contribution of the core excitations is given.}

\label{tab1}

\begin{tabular}{lcccccc}
\hline \hline
&&\multicolumn{1}{c}{~~~DFB~~~}
& \multicolumn{1}{c}{~~+RPA~~~}
& \multicolumn{1}{r}{~~+MBPT~~}
& \multicolumn{1}{r}{+``core''}
& \multicolumn{1}{r}{Total} \\
 & & $-0.418$  & $-0.058$ & $-0.033$  & +0.018  & $-0.491$ \\
\hline \hline
\end{tabular}
\end{table}

All mentioned corrections are presented in Table~\ref{tab1}.
Summing them up, we finally obtain:
\begin{equation}
    \langle 7s,F=5 || d_{\rm PNC} || 7s,F=4 \rangle
    = -0.49 \times 10^{-10} \, i \, \kappa \; {\rm a.u.}.
\label{s3}
\end{equation}
According to Table~\ref{tab1} the MBPT corrections to this
amplitude are relatively small. Therefore, we estimate the accuracy
of our result to be about few percent.

It is interesting to compare this amplitude to similar amplitudes
in K and Cs. The amplitude \eref{e5a} strongly depends on the
nuclear spin $I$, which is different for all alkalis. Therefore, it
is convenient to rewrite it in terms of the matrix element of the
electron operator $\bm{\sigma}=2\bm{s}$, as it was done in
Ref.~\cite{GEKM88}:
\begin{eqnarray}
  \langle 7s, \ F_f || d_{\rm PNC} || 7s, \ F_i \rangle
   &\equiv& i D \kappa
   \langle F_f||\sigma||F_i\rangle.
\label{s4}
\end{eqnarray}
In this form the parameter $D$ only weakly depends on the nuclear
spin $I$. Combining Eqs.~\eref{s3} and \eref{s4}, we get:
\begin{eqnarray}
    D = 10^{-12} \times \left\{
    \begin{array}{d@{\rm ~for~}rc}
     -0.07, & ^{39,41}{\rm K} & $\cite{GEKM88}$, \\
     -1.4,  & ^{133}{\rm Cs}  & $\cite{NK75t}$,  \\
    -11.0,  & ^{211}{\rm Fr}, &
    \end{array} \right.
\label{s5}
\end{eqnarray}
where we took into account the differences in definition of the
coupling constant $\kappa$ \cite{kap}.

One can see that the constant $D$ for Fr is almost an order of
magnitude larger, than for Cs. According to the
Refs.~\cite{DFS95,SJ00a}, the ratio of the spin-independent PNC
amplitudes for optical transitions $ns \rightarrow (n+1)s$ for Fr
$(n=7)$ and Cs $(n=6)$ is close to 20. That factor also accounts
for the 1.6 times difference of the weak charges $Q_W$ in the PNC
operator \eref{e1} for the two nuclei. Because the interaction of
the electron with the nuclear AM gives the main contribution to
the spin-dependent PNC amplitude, one can expect that the constant
$\kappa$ also grows as $A^{2/3}$ \cite{FK80}. That can account for
the extra factor $\sim 1.4$ for the amplitude \eref{s4} in Fr when
compared to that in Cs.

\paragraph{Acknowledgments}
   We are grateful to D.~DeMille, discussions with whom stimulated this
work.

\begin{thebibliography}{30}
\expandafter\ifx\csname bibnamefont\endcsname\relax
  \def\bibnamefont#1{#1}\fi
\expandafter\ifx\csname bibfnamefont\endcsname\relax
  \def\bibfnamefont#1{#1}\fi
\expandafter\ifx\csname url\endcsname\relax
  \def\url#1{\texttt{#1}}\fi
\expandafter\ifx\csname urlprefix\endcsname\relax\def\urlprefix{URL }\fi
\providecommand{\bibinfo}[2]{#2}
\providecommand{\eprint}[2][]{\url{#2}}

\bibitem{Khr91}
\bibinfo{author}{\bibfnamefont{I.~B.} \bibnamefont{Khriplovich}},
  \emph{\bibinfo{title}{Parity non-conservation in atomic phenomena}}
  (\bibinfo{publisher}{Gordon and Breach}, \bibinfo{address}{New York},
  \bibinfo{year}{1991}).

\bibitem{Zel57t}
\bibinfo{author}{\bibfnamefont{Y.~B.} \bibnamefont{Zel'dovich}},
  \bibinfo{journal}{ZhETF} \textbf{\bibinfo{volume}{33}},
  \bibinfo{pages}{1531} (\bibinfo{year}{1957})
 [\bibinfo{journal}{Sov. Phys.--JETP} \textbf{\bibinfo{volume}{6}},
  \bibinfo{pages}{1184} (\bibinfo{year}{1957})].

\bibitem{FK80}
\bibinfo{author}{\bibfnamefont{V.~V.} \bibnamefont{Flambaum}} \bibnamefont{and}
  \bibinfo{author}{\bibfnamefont{I.~B.} \bibnamefont{Khriplovich}},
  \bibinfo{journal}{ZhETF} \textbf{\bibinfo{volume}{79}},
  \bibinfo{pages}{1656} (\bibinfo{year}{1980})
 [\bibinfo{journal}{Sov. Phys.--JETP} \textbf{\bibinfo{volume}{52}},
  \bibinfo{pages}{835} (\bibinfo{year}{1980})].

\bibitem{FKS84a}
\bibinfo{author}{\bibfnamefont{V.~V.} \bibnamefont{Flambaum}},
  \bibinfo{author}{\bibfnamefont{I.~B.} \bibnamefont{Khriplovich}},
  \bibnamefont{and} \bibinfo{author}{\bibfnamefont{O.~P.}
  \bibnamefont{Sushkov}}, \bibinfo{journal}{Phys. Lett. B}
  \textbf{\bibinfo{volume}{146}}, \bibinfo{pages}{367} (\bibinfo{year}{1984}).

\bibitem{FK85a}
\bibinfo{author}{\bibfnamefont{V.~V.} \bibnamefont{Flambaum}} \bibnamefont{and}
  \bibinfo{author}{\bibfnamefont{I.~B.} \bibnamefont{Khriplovich}},
  \bibinfo{journal}{ZhETF} \textbf{\bibinfo{volume}{89}},
  \bibinfo{pages}{1505} (\bibinfo{year}{1985})
  [\bibinfo{journal}{Sov. Phys.--JETP} \textbf{\bibinfo{volume}{62}},
  \bibinfo{pages}{872} (\bibinfo{year}{1985})].

\bibitem{BP91b}
\bibinfo{author}{\bibfnamefont{C.}~\bibnamefont{Bouchiat}} \bibnamefont{and}
  \bibinfo{author}{\bibfnamefont{C.~A.} \bibnamefont{Piketty}},
  \bibinfo{journal}{Phys. Lett. B} \textbf{\bibinfo{volume}{269}},
  \bibinfo{pages}{195} (\bibinfo{year}{1991}).

\bibitem{WBC97}
\bibinfo{author}{\bibfnamefont{C.~S.} \bibnamefont{Wood}},
  \bibinfo{author}{\bibfnamefont{S.~C.} \bibnamefont{Bennett}},
  \bibinfo{author}{\bibfnamefont{D.}~\bibnamefont{Cho}},
  \bibinfo{author}{\bibfnamefont{B.~P.} \bibnamefont{Masterson}},
  \bibinfo{author}{\bibfnamefont{J.~L.} \bibnamefont{Roberts}},
  \bibinfo{author}{\bibfnamefont{C.~E.} \bibnamefont{Tanner}},
  \bibnamefont{and} \bibinfo{author}{\bibfnamefont{C.~E.}
  \bibnamefont{Wieman}}, \bibinfo{journal}{Science}
  \textbf{\bibinfo{volume}{275}}, \bibinfo{pages}{1759} (\bibinfo{year}{1997}).

\bibitem{FM97}
\bibinfo{author}{\bibfnamefont{V.~V.} \bibnamefont{Flambaum}} \bibnamefont{and}
  \bibinfo{author}{\bibfnamefont{D.~W.} \bibnamefont{Murray}},
  \bibinfo{journal}{Phys. Rev. C} \textbf{\bibinfo{volume}{56}},
  \bibinfo{pages}{1641} (\bibinfo{year}{1997}).

\bibitem{HW01}
\bibinfo{author}{\bibfnamefont{W.~C.} \bibnamefont{Haxton}} \bibnamefont{and}
  \bibinfo{author}{\bibfnamefont{C.~E.} \bibnamefont{Wieman}},
  \emph{\bibinfo{title}{Atomic parity nonconservation and nuclear anapole
  moments}} (\bibinfo{year}{2001}), \eprint{E-print: nucl-th/0104026}.

\bibitem{NK75t}
\bibinfo{author}{\bibfnamefont{V.~N.} \bibnamefont{Novikov}} \bibnamefont{and}
  \bibinfo{author}{\bibfnamefont{I.~B.} \bibnamefont{Khriplovich}},
  \bibinfo{journal}{Pis'ma v ZhETF} \textbf{\bibinfo{volume}{22}},
  \bibinfo{pages}{162} (\bibinfo{year}{1975})
 [\bibinfo{journal}{JETP Lett.} \textbf{\bibinfo{volume}{22}},
  \bibinfo{pages}{74} (\bibinfo{year}{1975})].

\bibitem{GEKM88}
\bibinfo{author}{\bibfnamefont{V.~G.} \bibnamefont{Gorshkov}},
  \bibinfo{author}{\bibfnamefont{V.~F.} \bibnamefont{Ezhov}},
  \bibinfo{author}{\bibfnamefont{M.~G.} \bibnamefont{Kozlov}},
  \bibnamefont{and} \bibinfo{author}{\bibfnamefont{A.~I.}
  \bibnamefont{Mikhailov}},
  \bibinfo{journal}{Yad. Fiz.} \textbf{\bibinfo{volume}{48}},
  \bibinfo{pages}{1363} (\bibinfo{year}{1988})
 [\bibinfo{journal}{Sov. J. Nucl. Phys.} \textbf{\bibinfo{volume}{48}},
  \bibinfo{pages}{867} (\bibinfo{year}{1988})].

\bibitem{Fla88}
\bibinfo{author}{\bibfnamefont{V.~V.} \bibnamefont{Flambaum}}
  (\bibinfo{year}{1988}), \bibinfo{note}{in the strong dc magnetic field the
  ${M1}$ amplitude for the transition between Zeeman levels can be suppressed
  (unpublished)}.

\bibitem{Beh93}
\bibinfo{author}{\bibfnamefont{J.~A.} \bibnamefont{Behr}},
  \bibinfo{author}{\bibfnamefont{S.~B.} \bibnamefont{Cahn}},
  \bibinfo{author}{\bibfnamefont{S.~B.} \bibnamefont{Dutta}},
  \bibinfo{author}{\bibfnamefont{A.}~\bibnamefont{Gorlitz}}, \emph{et~al.},
  \bibinfo{journal}{Hyperfine Interact.} \textbf{\bibinfo{volume}{81}},
  \bibinfo{pages}{197} (\bibinfo{year}{1993}).

\bibitem{Oro97}
\bibinfo{author}{\bibfnamefont{L.~A.} \bibnamefont{Orozco}} \emph{et~al.},
  \bibinfo{journal}{AIP Conf. Proc.} \textbf{\bibinfo{volume}{400}},
  \bibinfo{pages}{107} (\bibinfo{year}{1997}).

\bibitem{GOP99}
\bibinfo{author}{\bibfnamefont{J.~S.} \bibnamefont{Grossman}},
  \bibinfo{author}{\bibfnamefont{L.~A.} \bibnamefont{Orozco}},
  \bibinfo{author}{\bibfnamefont{M.~R.} \bibnamefont{Pearson}},
  \bibinfo{author}{\bibfnamefont{J.~E.} \bibnamefont{Simsarian}},
  \bibinfo{author}{\bibfnamefont{G.~D.} \bibnamefont{Sprouse}},
  \bibnamefont{and} \bibinfo{author}{\bibfnamefont{W.~Z.} \bibnamefont{Zhao}},
  \bibinfo{journal}{Phys. Rev. Lett.} \textbf{\bibinfo{volume}{83}},
  \bibinfo{pages}{935} (\bibinfo{year}{1999}).

\bibitem{Bv83}
\bibinfo{author}{\bibfnamefont{P.}~\bibnamefont{Bogdanovich}} \bibnamefont{and}
  \bibinfo{author}{\bibfnamefont{G.}~\bibnamefont{\v{Z}ukauskas}},
  \bibinfo{journal}{Sov. Phys. Collection}
  \textbf{\bibinfo{volume}{23}}(\bibinfo{number}{5}), \bibinfo{pages}{18}
  (\bibinfo{year}{1983}).

\bibitem{Bog91}
\bibinfo{author}{\bibfnamefont{P.}~\bibnamefont{Bogdanovich}},
  \bibinfo{journal}{Lithuanian Physics Journal}
  \textbf{\bibinfo{volume}{31}}(\bibinfo{number}{2}), \bibinfo{pages}{79}
  (\bibinfo{year}{1991}).

\bibitem{KPF96}
\bibinfo{author}{\bibfnamefont{M.~G.} \bibnamefont{Kozlov}},
  \bibinfo{author}{\bibfnamefont{S.~G.} \bibnamefont{Porsev}},
  \bibnamefont{and} \bibinfo{author}{\bibfnamefont{V.~V.}
  \bibnamefont{Flambaum}}, \bibinfo{journal}{J. Phys. B}
  \textbf{\bibinfo{volume}{29}}, \bibinfo{pages}{689} (\bibinfo{year}{1996}).

\bibitem{KP97}
\bibinfo{author}{\bibfnamefont{M.~G.} \bibnamefont{Kozlov}} \bibnamefont{and}
  \bibinfo{author}{\bibfnamefont{S.~G.} \bibnamefont{Porsev}},
  \bibinfo{journal}{ZhETF} \textbf{\bibinfo{volume}{111}},
  \bibinfo{pages}{831} (\bibinfo{year}{1997})
 [\bibinfo{journal}{Sov. Phys.--JETP} \textbf{\bibinfo{volume}{84}},
  \bibinfo{pages}{461} (\bibinfo{year}{1997})].

\bibitem{Ste50}
\bibinfo{author}{\bibfnamefont{R.~M.} \bibnamefont{Sternheimer}},
  \bibinfo{journal}{Phys. Rev.} \textbf{\bibinfo{volume}{80}},
  \bibinfo{pages}{102} (\bibinfo{year}{1950}).

\bibitem{DL55}
\bibinfo{author}{\bibfnamefont{A.}~\bibnamefont{Dalgarno}} \bibnamefont{and}
  \bibinfo{author}{\bibfnamefont{J.~T.} \bibnamefont{Lewis}},
  \bibinfo{journal}{Proc. Roy. Soc.} \textbf{\bibinfo{volume}{223}},
  \bibinfo{pages}{70} (\bibinfo{year}{1955}).

\bibitem{KP99}
\bibinfo{author}{\bibfnamefont{M.~G.} \bibnamefont{Kozlov}} \bibnamefont{and}
  \bibinfo{author}{\bibfnamefont{S.~G.} \bibnamefont{Porsev}},
  \bibinfo{journal}{Eur. Phys. J. D} \textbf{\bibinfo{volume}{5}},
  \bibinfo{pages}{59} (\bibinfo{year}{1999}).

\bibitem{DFS95}
\bibinfo{author}{\bibfnamefont{V.~A.} \bibnamefont{Dzuba}},
  \bibinfo{author}{\bibfnamefont{V.~V.} \bibnamefont{Flambaum}},
  \bibnamefont{and} \bibinfo{author}{\bibfnamefont{O.~P.}
  \bibnamefont{Sushkov}}, \bibinfo{journal}{Phys. Rev. A}
  \textbf{\bibinfo{volume}{51}}, \bibinfo{pages}{3454} (\bibinfo{year}{1995}).

\bibitem{SJ00a}
\bibinfo{author}{\bibfnamefont{M.~S.} \bibnamefont{Safronova}}
  \bibnamefont{and} \bibinfo{author}{\bibfnamefont{W.~R.}
  \bibnamefont{Johnson}}, \bibinfo{journal}{Phys. Rev. A}
  \textbf{\bibinfo{volume}{62}}, \bibinfo{pages}{022112}
  (\bibinfo{year}{2000}).

\bibitem{kap}
\bibinfo{note}{It should be mentioned that several definitions of the
coupling constant $\kappa$ are used in the literature, and one has
to take it into account when comparing different results}.
\end{thebibliography}


\end{document}